\documentclass[%
reprint,
twocolumn,
superscriptaddress,
amsmath,amssymb,
aps,
floatfix,
rc ]{revtex4-2}

\usepackage{graphicx}
\usepackage{dcolumn}
\usepackage{bm}
\usepackage{nicematrix}

\usepackage[hidelinks]{hyperref}

\usepackage{float}
\usepackage{color}
\usepackage{soul}
\usepackage{braket}
\usepackage{leftindex}
\usepackage{booktabs}
\usepackage[T1]{fontenc}
\usepackage{siunitx}

\begin{document}

\preprint{APS/123-QED}

\title{Entanglement of a nuclear spin qubit register in silicon photonics}

\author{Hanbin Song}
\affiliation{Materials Sciences Division, Lawrence Berkeley National Laboratory, Berkeley, California 94720, USA}
\affiliation{
Department of Materials Science and Engineering, University of California, Berkeley, Berkeley, California 94720, USA
}
\author{Xueyue Zhang}
\email{Current address: Department of Applied Physics and Applied Mathematics, Columbia University, New York, New York 10027, USA}
\affiliation{Department of Electrical Engineering and Computer Sciences, University of California, Berkeley, Berkeley, California 94720, USA}
\affiliation{Department of Physics, University of California, Berkeley, Berkeley, California 94720, USA}

\author{Lukasz Komza}
\affiliation{Department of Physics, University of California, Berkeley, Berkeley, California 94720, USA}
\affiliation{Materials Sciences Division, Lawrence Berkeley National Laboratory, Berkeley, California 94720, USA}

\author{Niccolo Fiaschi}
\affiliation{Accelerator Technology and Applied Physics Division, Lawrence Berkeley National Laboratory, Berkeley, CA, USA}
\affiliation{Department of Electrical Engineering and Computer Sciences, University of California, Berkeley, Berkeley, California 94720, USA}
\affiliation{Department of Physics, University of California, Berkeley, Berkeley, California 94720, USA}
\author{Yihuang Xiong} 
\affiliation{Thayer School of Engineering, Dartmouth College, Hanover, New Hampshire 03755, USA}
\author{Yiyang Zhi}
\affiliation{Department of Electrical Engineering and Computer Sciences, University of California, Berkeley, Berkeley, California 94720, USA}

\author{Scott~Dhuey}
\affiliation{Molecular Foundry, Lawrence Berkeley National Laboratory, Berkeley, California 94720, USA}

\author{Adam~Schwartzberg}
\affiliation{Molecular Foundry, Lawrence Berkeley National Laboratory, Berkeley, California 94720, USA}

\author{Thomas~Schenkel}
\affiliation{Accelerator Technology and Applied Physics Division, Lawrence Berkeley National Laboratory, Berkeley, CA, USA}

\author{Geoffroy Hautier} 
\affiliation{Thayer School of Engineering, Dartmouth College, Hanover, New Hampshire 03755, USA}
\affiliation{Department of Materials Science and NanoEngineering, Rice University, Houston, TX, 77005}
\affiliation{Rice Advanced Materials Institute, Rice University, Houston, TX, 77005}

\author{Zi-Huai Zhang}
\affiliation{Department of Electrical Engineering and Computer Sciences, University of California, Berkeley, Berkeley, California 94720, USA}
\affiliation{Materials Sciences Division, Lawrence Berkeley National Laboratory, Berkeley, California 94720, USA}
\affiliation{Department of Physics, University of California, Berkeley, Berkeley, California 94720, USA}

\author{Alp Sipahigil}
\email{Corresponding author: alp@berkeley.edu}
\affiliation{Department of Electrical Engineering and Computer Sciences, University of California, Berkeley, Berkeley, California 94720, USA}
\affiliation{Materials Sciences Division, Lawrence Berkeley National Laboratory, Berkeley, California 94720, USA}
\affiliation{Department of Physics, University of California, Berkeley, Berkeley, California 94720, USA}

\date{\today}

\begin{abstract}
Color centers provide an optical interface to quantum registers based on electron and nuclear spin qubits in solids. The T center in silicon is an emerging spin-photon interface that combines telecom O-band optical transitions and an electron spin in a scalable photonics platform. In this work, we demonstrate the initialization, coherent control, and state readout of a three-qubit register based on the electron spin of a T center coupled to a hydrogen and a silicon nuclear spin. The spin register exhibits spin echo coherence times of $0.41(2)$~ms for the electron spin, $112(12)$~ms for the hydrogen nuclear spin, and $67(7)$~ms for the silicon nuclear spin. We use nuclear-nuclear two-qubit gates to generate entanglement between the two nuclear spins with a fidelity of $F=0.77(3)$ and a coherence time of $T^*_2=2.60(8)$~ms. Our results show that a T center in silicon photonics can realize a multi-qubit register with an optical interface for quantum communication.
\end{abstract}

\maketitle

Spin-photon interfaces based on color centers in solids are well-suited as optically interconnected multi-qubit registers for quantum information processing~\cite{awschalom_quantum_2018, gao_coherent_2015, atature_material_2018}. Randomly placed nuclear spins surrounding color centers can form a multi-qubit register, which couples to the optically addressable electron spin of the color center via hyperfine interaction~\cite{taminiau_detection_2012, bradley_ten-qubit_2019}. In host materials with low nuclear spin concentrations~\cite{Ferrenti2020}, the electron and nuclear spins display long coherence times and are ideal platforms for storing quantum information~\cite{Onizhuk2025Gali}. In particular, coherent nuclear spins with an optical interface via the electron spin can serve as quantum memories for long-distance quantum communication~\cite{hensen_loophole-free_2015, knaut_entanglement_2024, ruskuc_multiplexed_2025}.

Color centers in silicon are emerging spin-photon interfaces~\cite{bergeron_silicon-integrated_2020,higginbottom_optical_2022} that show promise for scalability due to silicon's unique capabilities for scalable, high-performance electronic-photonic integration using advanced semiconductor foundry nodes~\cite{psiquantum_manufacturable_2025}. Among the silicon color centers studied so far~\cite{Durand}, the T center is the only system combining a telecom O-band emission with a coherent electron spin in the ground state~\cite{bergeron_silicon-integrated_2020}. T centers integrated into photonic devices allow an efficient optical interface for this defect~\cite{islam_cavity-enhanced_2024, johnston_cavity-coupled_2024, komza_multiplexed_2025,inc_distributed_2024}. Coherent control of the electron and hydrogen nuclear spin associate with a single T center has been demonstrated in isotopically purified substrates~\cite{inc_distributed_2024}. However, the possibility of building a coherent multi-qubit register based on T centers coupled to the $^{29}$Si nuclear spin bath in natural silicon has not been explored.

In this work, we report a three-qubit register based on a single T center in a silicon photonic waveguide. The three-qubit register consists of the T center electron spin, the T center hydrogen nuclear spin, and a nearby $^{29}$Si nuclear spin. We demonstrate coherent control of all three qubits and multi-qubit controlled-NOT gates. The three qubits show coherence times of $T_\mathrm{{2, e}}^\mathrm{{echo}}=411(15)~\mathrm{\mu s}$, $T_\mathrm{{2, H}}^\mathrm{{echo}}=110(10)~\mathrm{ms}$, and $T_\mathrm{{2,{Si}}}^\mathrm{{echo}}=70(7)~\mathrm{ms}$. We create entanglement between the two nuclear spins with a fidelity of 77(3)\% and a coherence time of $2.60(8)$~ms. The observed coherence times for single qubits and entangled states demonstrate that the T center can be used to realize a multi-qubit quantum memory.

\begin{figure}[h]
    \includegraphics[]{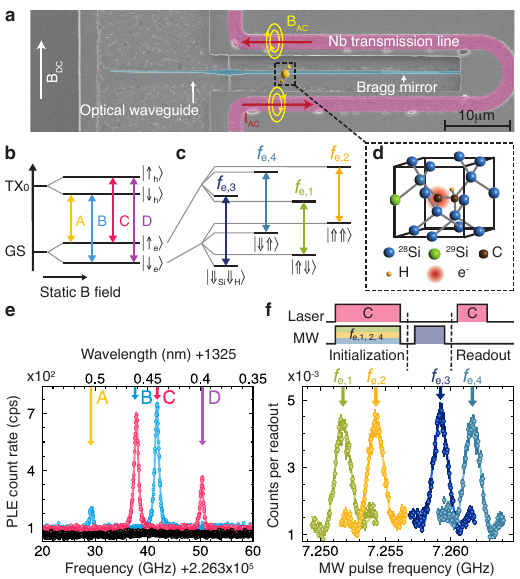}
    \caption{\textbf{a,} False-colored scanning electron microscope image of the optical waveguide (silicon, blue) and the superconducting transmission line (niobium, pink). A magnet provides a static in-plane magnetic field $\vec{\mathrm{B}}_\mathrm{DC}$. \textbf{b, c,} Energy level diagram of a T center in a magnetic field. The electron spin is coupled to H and Si nuclear spins via hyperfine interactions. \textbf{d,} Atomic structure of a T center with a neighboring $^{29}$Si atom. \textbf{e,} Two-tone photoluminescence excitation (PLE) measurement of the electron spin-dependent optical transitions. The blue (red) trace shows the PLE spectrum with a pump frequency indicated by the blue (red) arrow. Transitions B and C exhibit a linewidth of 0.76~GHz and are detuned by 3.35 GHz. Black: single laser PLE spectrum. \textbf{f,} Upper: pulse sequence for optically detected magnetic resonance (ODMR) of the $f_{e,3}$ transition. The sequence initializes the system in the  $\lvert\downarrow_\mathrm{e} \Downarrow_\mathrm{Si} \Downarrow_\mathrm{H} \rangle$ state. Driving optical transition C polarizes the electron spin to the $\lvert\downarrow_e\rangle$ manifold and driving $f_{e: 1,2,4}$ depletes the population in $\lvert\Uparrow\Downarrow\rangle, \lvert\Uparrow\Uparrow\rangle$, and $\lvert\Downarrow\Uparrow\rangle$ nuclear states. Lower: Pulsed-ODMR spectra of nuclear spin-dependent electron spin resonances for different nuclear initialization states.}
    \label{fig:fig1}
\end{figure}
\noindent{\textbf{T center in a photonic waveguide.} }The T center in silicon is a point defect that contains two carbon atoms, one hydrogen atom, and an unpaired electron in a silicon substitutional site (Fig.~\ref{fig:fig1}\textbf{d})~\cite{irion_defect_1985, safonov_interstitial-carbon_1996}. The T center ground state (GS) comprises an electron spin ($S=\frac{1}{2}$), while its first excited state ($\text{TX}_0$) is a defect-bound exciton state with an unpaired hole spin ($S=\frac{1}{2}$)~\cite{safonov_interstitial-carbon_1996}. The optical emission from $\text{TX}_0$ has a zero-phonon line (ZPL) around 1326 nm in the telecom O-band with a Debye-Waller factor of 0.23~\cite{bergeron_silicon-integrated_2020}. Under a static magnetic field, the Zeeman splitting results in spectrally resolved, spin-dependent optical transitions (Fig.~\ref{fig:fig1}\textbf{b}). The electron spin g-factor is isotropic ($g_e = 2.005$), while the hole spin g-factor ($g_h$) depends strongly on the relative alignment between magnetic field direction and T center orientation~\cite{clear_optical-transition_2024}. Each T center intrinsically contains a hydrogen nuclear spin ($I=\frac{1}{2}$). Additionally, the silicon lattice contains 4.67\% of $^{29}$Si with nuclear spin $I=\frac{1}{2}$. Our first-principles calculations show that 46 lattice sites surrounding the T center can have a hyperfine coupling strength above 1~MHz (Appendix \ref{SI: silicon hyperfine}). Consequently, the T center electron is likely to strongly couple to nearby $^{29}$Si nuclei, forming a quantum register with one electron spin and hyperfine coupled nuclear spins. 

We generate T centers by implanting ions ($\left[^{12}\text{C}\right]= 1\times 10^{12}~\mathrm{cm^{-2}}$, $\left[\text{H}\right]=7\times10^{12}~\mathrm{cm^{-2}}$) into a high-resistivity silicon-on-insulator (SOI) wafer followed by rapid thermal annealing. The low implantation dose~\cite{MacQuarrie2021} results in spatially resolvable T centers under free space excitation. We integrate T centers in single-mode photonic waveguides. Each photonic waveguide is terminated with a Bragg reflector and is adiabatically tapered to mode match to a lensed fiber for efficient single-sided photon collection (Fig.~\ref{fig:fig1}\textbf{a}). A superconducting transmission line is patterned adjacent to the photonic waveguides for spin control. The device is cooled down to 3.4~K in a closed-cycle cryostat. Device fabrication and measurement setup details are available in Appendix~\ref{SI: set up},~\ref{SI: fabrication}.

We locate T centers by scanning an off-resonant laser (635~nm) across a photonic waveguide and detecting T center zero-phonon line (ZPL) emission into the waveguide centered at 1325.4~nm. To resolve spin-dependent transitions of a T center, we perform resonant photoluminescence excitation (PLE) through free space excitation and collect the T center fluorescence in the phonon sideband (PSB) via the waveguide. Due to the presence of an external magnetic field, excitation with a single resonant laser optically pumps the electron spin to a dark state, preventing the observation of steady-state PLE signal (Fig.~\ref{fig:fig1}\textbf{e}, black line). We perform two-tone laser spectroscopy by scanning two lasers around the ZPL. The T center fluorescence shows up only if the two lasers excite transitions from different electron spin states simultaneously (Fig.~\ref{fig:fig1}\textbf{e}). Using the splitting of optical transitions and assuming $g_e = 2.005$~\cite{clear_optical-transition_2024}, we infer the electron spin resonance (ESR) frequency to be $f_e= 7.3~\text{GHz}$ and the local magnetic field to be \mbox{B = 0.26~T}.

\noindent{\textbf{Electron spin control and coherence.} }We use optically detected magnetic resonance (ODMR) to study the electron spin coherence and hyperfine structure. In the presence of strongly coupled nuclear spins, the electron spin resonances split into spectrally resolved, nuclear spin-dependent transitions which cannot be simultaneously addressed with a single microwave tone. We use pulsed ODMR measurements involving multi-tone microwave pulses (see Appendix \ref{SI:search ESR}) to identify the nuclear spin-dependent electron spin resonances and to initialize the electron-nuclear spin register. The pulsed measurements in Fig.~\ref{fig:fig1}\textbf{f} reveal a hyperfine structure consisting of four electron spin resonances at frequencies $f_{e,1}, f_{e,2}, f_{e,3},f_{e,4}$, consistent with having a T center strongly coupled to nuclear spin of the hydrogen and a nearby $^{29}$Si (Fig.~\ref{fig:fig1}\textbf{c}). These transitions are spectrally resolved, enabling controlled rotation gates (e.g. $\mathrm{C_{H,~Si}NOT_{e}}$) where the electron spin rotations are conditioned on the nuclear spin states. Due to perpendicular components of the hyperfine interaction, we also observe nuclear spin non-conserving electron spin transitions (see Appendix \ref{SI:cross_transitions}). 

 \begin{figure}[h]
    \centering
    \includegraphics[]{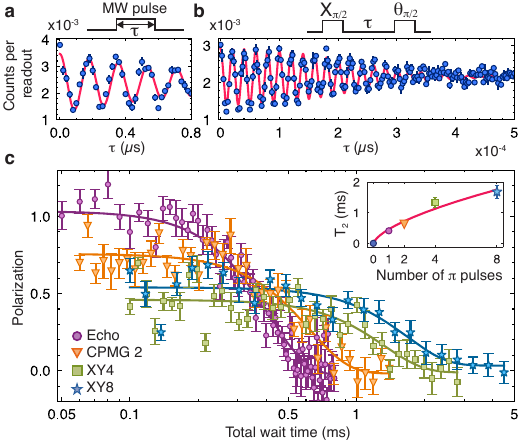}
    \caption{\textbf{T center electron spin coherence.} \textbf{a,} Rabi oscillation of the electron spin with \mbox{$t_{\pi}=90$~ns}. \textbf{b,} Ramsey fringes of the electron spin show \mbox{$T_{2, e}^*=2.7(1)~\mu$s}. A 5~MHz virtual detuning ($\theta(\tau)$) is added to the second $\pi/2$ pulse phase. \textbf{c,} Electron spin coherence time using different dynamical decoupling sequences: \mbox{$T_{2, e}^{\text{echo}}=411(15)~\mu$s}, \mbox{$T_{2, e}^{\text{CPMG2}} = 623(43)~\mu$s}, \mbox{$T_{2, e}^{\text{XY4}}=1.33(17)$~ms}, and \mbox{$T_{2, e}^{\text{XY8}}=1.68(22)$~ms}. The polarization is normalized using the Hahn-echo measurement contrast. We estimate a $\pi$ pulse fidelity of 92(2)\% based on polarization decay with increasing number of refocusing pulses (see Appendix \ref{SI:pulse fidelity}). Inset: coherence time versus number of refocusing pulses. The system is initialized in $\ket{\downarrow_{e} \Downarrow_{\text{Si}} \Downarrow_{\text{H}}}$ before all measurements. MW pulses resonant with the nuclear spin conserving transition frequency ($f_{e, 3}$) are used to drive the electron spin. A stretching factor of 2.5 is used when fitting coherence decays.}
    \label{fig:fig2}
\end{figure}

We initialize the nuclear spins using a multi-tone pulse sequence (Fig.~\ref{fig:fig1}\textbf{f}) to allow high-fidelity control of the electron spin with resonant pulses (Fig.~\ref{fig:fig2}\textbf{a}). The electron spin shows a Ramsey coherence time of $T_{2, e}^*~=~2.7(1)~\mathrm{\mu s}$ (Fig.~\ref{fig:fig2}\textbf{b}), which is likely limited by the $^{29}$Si nuclear spin bath~\cite{inc_distributed_2024}. Using a Hahn-echo sequence suppresses the slow magnetic noise and extends the spin coherence time to $T_{2, e}^{\text{echo}}=411(15)~\mu $s (Fig.~\ref{fig:fig2}\textbf{c}). Interestingly, our observed spin coherence time $T_{2, e}^{\text{echo}}$ for a device-integrated T center in natural silicon is longer than that in isotopically-enriched silicon ($T_{2, e}^{\text{echo}}=270(10)$~$\mu s$) in Ref.~\cite{inc_distributed_2024}. Further studies are needed to understand the discrepancy and sources limiting $T_{2,e}$ of T centers in different devices. Applying dynamical decoupling with an XY8 sequence further extends the coherence time to $1.68(22)$~ms (Fig.~\ref{fig:fig2}\textbf{c}). For all dynamical decoupling measurements, we change the phase of the last $\pi/2$ pulse of the sequences to project the spin to either up or down states, and read out using the same laser to calculate polarization. We fit the coherence time as a function of the number of refocusing pulses ($N$) using $T_{\text{coh}}(N)\propto \alpha N^n$ and find a stretching factor of $n=0.62(16)$, in agreement with a noise spectrum from a Lorentzian bath where $n=2/3$ \cite{de_lange_universal_2010}. We do not observe any depolarization of the electron spin within $100~$ms in $T_1$ measurement (Fig.~\ref{fig:SI_electron_T1})~\cite{higginbottom_optical_2022}. 

\begin{figure*}
    \includegraphics{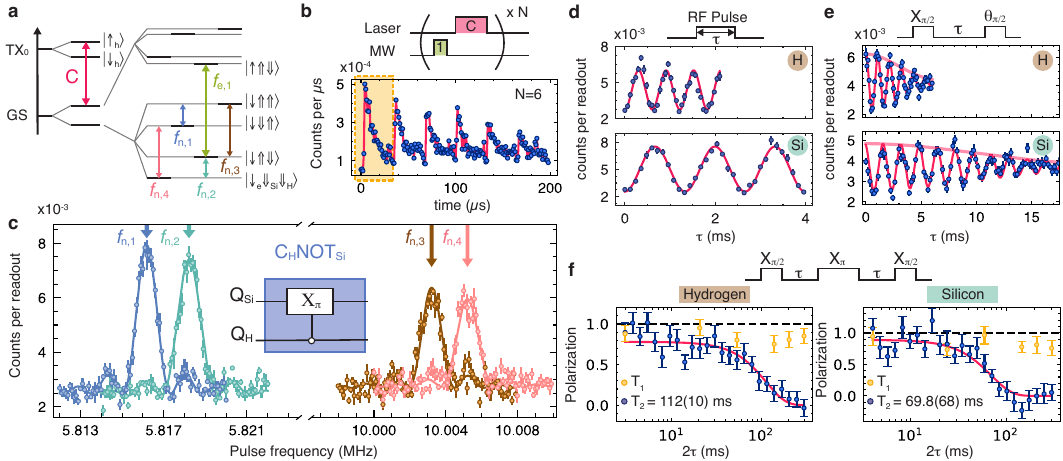}
    \caption{\textbf{Nuclear spin resonances and coherences.} \textbf{a,} A zoomed-in view of the hyperfine structure of the T center under a magnetic field. \textbf{b,} Upper: pulse sequence to readout population in the $\lvert\downarrow_e\Uparrow_\text{Si}\Downarrow_\text{H}\rangle$ state. A readout unit contains a C$_{\text{nn}}$NOT$_\text{e}$ pulse and a laser pulse that probes the population in the $\lvert\uparrow_{e}\rangle$ manifold. Lower: A histogram of repetitive nuclear spin readout with $\text{N}=6$. The dashed box indicates a readout cycle. Electron spin flips during optical excitation results in the decay of fluorescence in each readout cycle. Nuclear spin flips during optical excitation leads to the decay of total fluorescence with increasing N. \textbf{c,} Nuclear spin resonances in the $\lvert\downarrow_e\rangle$ manifold. Inset: state-selective $\pi$ pulses constitute $\text{C}_n\text{NOT}_n$ gates between the two nuclear spins. \textbf{d,} Rabi oscillation of the H and Si nuclear spins driven with resonant RF pulses. \textbf{e,} Ramsey coherence measurement of the H and Si nuclear spins, $T_{2, H}^*=4.0(2)$~ms, and $T_{2,\text{Si}}^*=13.9(8)$~ms. \textbf{f,} Hahn-echo decay traces of the H and Si nuclear spins (navy). Measurements of nuclear spin lifetime ($T_1$) show no observable decay up to 300~ms (yellow).}
    \label{fig:fig3}
\end{figure*}
\noindent{\textbf{Nuclear spin control and coherence.} } To probe nuclear spin coherences, we first study the hyperfine structure of the T center (Fig.~\ref{fig:fig3}\textbf{a}) using pulsed-ODMR. During pulsed-ODMR, we read out nuclear spin population using a controlled NOT ($\mathrm{C_{H,~Si}NOT_{e}}$) gate that selectively maps the nuclear spin state to the electron spin (Fig.~\ref{fig:fig3}\textbf{b}) which is optically read out. Fig.~\ref{fig:fig3}\textbf{b} shows repeated applications of the readout sequence and the resulting fluorescence signal. The nuclear spin state is partially preserved during optical excitation for electron spin state readout. Based on the spin cyclicities under optical excitation, we use two repetitive readout cycles to increase the signal-to-noise ratio \cite{jiang_repetitive_2009, neumann_single-shot_2010}. 

We measure the nuclear magnetic resonances in the $\lvert\downarrow_e\rangle$ manifold by sweeping a radio-frequency pulse. We observe four nuclear spin resonances consisting of two groups centered at $5.817~$MHz and $10.004~$MHz that correspond to silicon and hydrogen nuclear spins, respectively (Fig.~\ref{fig:fig3}\textbf{c}). Within each group, the transition frequency of the nuclear spin (e.g. $f_{n,1}$ and $f_{n,2}$) shifts by 2~kHz conditioned on the initialization state of the other nuclear spin. The $2$~kHz effective longitudinal coupling between the two nuclear spins results in frequency resolvable state-dependent transitions, enabling  $\mathrm{C_{H}NOT_{Si}}$ and $\mathrm{C_{Si}NOT_{H}}$ gates. The longitudinal coupling originates from a combination of direct nuclear dipolar interaction and electron spin mediated nuclear spin interactions~\cite{vandeStolpe2024, shi_sensing_2014, bartling_entanglement_2022}.

With the nuclear resonances known, we calculate the nuclear gyromagnetic ratio to confirm the assignment of the nuclear resonances to the silicon and hydrogen spin. For the hydrogen nuclear spin, the transition resonance is at $f_{n, 3}=10.005$~MHz for the $\lvert\downarrow_{\text{e}} \Downarrow_{\text{Si}}\rangle$ state, and 12.141~MHz for the $\lvert\uparrow_{\text{e}} \Downarrow_{\text{Si}} \rangle$ state (Fig.~\ref{fig:up_e_H}\textbf{a}). We calculate the nuclear Zeeman splitting to be 11.073~MHz and the hyperfine longitudinal coupling to be 1.07~MHz. Using the strength of the local magnetic field obtained from optical splittings, the experimental gyromagnetic ratio of the hydrogen nuclear spin is $\gamma/(2\pi)=42.49$~MHz/T, which closely matches with the literature value of $\gamma_\text{H}/(2\pi)=42.577$~MHz/T. For comparison, the silicon gyromagnetic ratio is $\gamma_{^{29}\text{Si}}/(2\pi)=-8.465$~MHz/T. We identify the $5.8$~MHz resonances with the silicon nuclear spin, which has the highest abundance in the host lattice (see Appendix~\ref{SI:Up_e NMR}). 

We demonstrate coherent control of hydrogen and silicon nuclear spins using resonant RF pulses (Fig.~\ref{fig:fig3}\textbf{d}). At the same RF power, the hydrogen nuclear spin exhibits a higher Rabi frequency than silicon, consistent with hydrogen's larger gyromagnetic ratio. The larger gyromagnetic ratio also implies stronger coupling to the noisy spin bath, which explains the shorter Ramsey coherence time for hydrogen $T_{2, \text{H}}^*=4.0(2)$~ms compared to silicon $T_{2, \text{Si}}^*=13.9(8)$~ms (Fig.~\ref{fig:fig3}\textbf{e}). Noise cancellation using a Hahn-echo sequence extends the nuclear coherence times to $T_{2, \text{H}}^{\text{echo}}=112(10)$~ms, and $T_{2, \text{Si}}^{\text{echo}}=70(7)$~ms (Fig.~\ref{fig:fig3}\textbf{f}). Our observation of $T_{2, \text{H}}^{\text{echo}}>T_{2, \text{Si}}^{\text{echo}}$ suggests that further studies \cite{taminiau_detection_2012} are needed to understand the different noise sources that the two nuclear spins experience. Finally, nuclear spin lifetime ($T_1$) measurements show no population decay up to 300~ms, indicating that the Hahn-echo coherence time is not lifetime-limited. 


\begin{figure}
    \centering
    \includegraphics[]{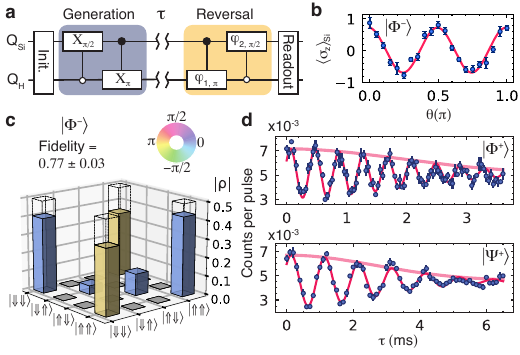}
    \caption{\textbf{Nuclear spin entanglement.} \textbf{a,} Circuit diagram for Bell state preparation and phase reversal tomography. \textbf{b,} Si nuclear spin expectation value as a function of the phase of reversal pulses. \textbf{c,} Reconstructed density matrix of the $\lvert\Phi^-\rangle$ Bell state. Matrix elements that cannot be extracted by phase reversal tomography are marked in gray. The color indicates the phase of the off-diagonal terms. Dashed columns show the density matrix of an ideal $\lvert\Phi^-\rangle$ Bell state. \textbf{d,} Ramsey fringes for $\lvert\Phi^+\rangle$: $T_{2}^* = 2.60(8)$~ms, and $\lvert\Psi^+\rangle$: $T_{2}^*=3.75(8)$~ms.}
    \label{fig:fig4}
\end{figure}

\noindent{\textbf{Nuclear spin entanglement.}} We use controlled-NOT gates between the H and Si nuclear spins to entangle the two nuclear spins. An example sequence of entanglement generation between H and Si in the $\lvert\downarrow_e\rangle$ manifold is shown in Fig.~\ref{fig:fig4}\textbf{a}. The system is first initialized in $\lvert\downarrow_{\text{e}} \Uparrow_{\text{Si}} \Uparrow_{\text{H}}\rangle$ using the sequence shown in Fig.~\ref{fig:SI_tomography}\textbf{a}. Then, two controlled rotation gates (pulses at frequencies $f_{n,1}$ and $f_{n,4}$) are applied to create the Bell state \mbox{$\lvert\Phi^-\rangle=\frac{1}{\sqrt{2}}(\lvert\Uparrow \Uparrow \rangle - \lvert\Downarrow \Downarrow \rangle)$}. To quantify the fidelity of the Bell state, we perform phase reversal tomography \cite{mehring_entanglement_2003, stemp_tomography_2024}. The two controlled rotation gates used for Bell state generation are applied in reverse order with the rotation axes offset by azimuthal angles ($\varphi_1=\theta,~\varphi_2=3\theta$) via microwave pulse phases. This tomography scheme maps the non-zero off-diagonal density matrix elements of a Bell state to the diagonal matrix elements of the final state, which can be measured by the expectation value of the nuclear spin $\langle \sigma_z \rangle_{\text{Si}}$ (Fig.~\ref{fig:fig4}\textbf{b}). Varying $\theta$ creates an oscillation in $\langle \sigma_z \rangle_{\text{Si}}$, whose phase and amplitude encode the non-zero off-diagonal matrix elements of the Bell state density matrix. Together with diagonal matrix elements obtained from population measurement directly after Bell state preparation (Fig.~\ref{fig:SI_tomography}\textbf{d}), we reconstruct key elements of the density matrix and calculate the fidelity to be $F_{\lvert\Phi^-\rangle} = 77(3)$\% (Fig.~\ref{fig:fig4}\textbf{c}). The fidelity is limited by imperfect state initialization and pulse errors (see Appendix \ref{SI:Tomography}). To probe coherence of the entangled states, we measure the dephasing times of $|\Phi^+\rangle=\frac{1}{\sqrt{2}}(\lvert\Uparrow \Uparrow \rangle + \lvert\Downarrow \Downarrow \rangle)$ and $|\Psi^+\rangle=\frac{1}{\sqrt{2}}(\lvert\Uparrow \Downarrow\rangle + \lvert\Downarrow \Uparrow \rangle)$ by adding a free evolution time ($\tau$) between state generation and reversal sequences. For the two reversal pulses, the azimuthal angles are set to $\varphi_1=0,~\varphi_2=\omega\tau$ via microwave pulse phases, where $\omega$ is a virtual detuning. 
By fitting the decay of $\langle \sigma_z \rangle_{\text{Si}}$, we measure $T_{2, \lvert\Phi^+\rangle}^* = 2.60(8)$~ms, and $T_{2, \lvert\Psi^+\rangle}^*=3.75(8)$~ms (Fig.~\ref{fig:fig4}\textbf{d}). The odd-parity $\lvert\Psi^+\rangle$ state has a longer dephasing time than the even-parity $\lvert\Phi^+\rangle$ state, suggesting the noise sources on the two nuclear spins are partially correlated (see Appendix~\ref{SI:Bell_coherence}). 

\noindent{\textbf{Discussion.} }We demonstrated the initialization, coherent control, and readout of a three-qubit register in a silicon photonics platform. Our results show that electron-nuclear and nuclear-nuclear two-qubit gates are readily available for T centers in natural silicon, and can be used to generate entangled states within the spin-qubit register. The nuclear spins show echo coherence times of $\sim 100$~ms and the nuclear spin Bell states show Ramsey coherence times of $\sim 3$~ms without any isotopic purification.

Several factors contribute to errors in single- and two-qubit gate operations within the nuclear register and state readout.  The 2~kHz conditional frequency shift between the nuclear spins  poses a challenge for performing high-fidelity nuclear single-qubit gates. A single-qubit gate  (unconditional on the other nuclear spin) requires a large driving field and Rabi frequency ($\gg2$~kHz), but is not achievable using our transmission line without breaking superconductivity and causing heating (see Appendix~\ref{SI:freq_jump}). Using superconducting films with higher critical currents, together with careful microwave and thermal engineering, may enable high-fidelity single-qubit gates. The nuclear two-qubit gates rely on selective excitation of two transitions split by 2 kHz. Similarly, electron-nuclear two qubit gates rely on selective excitations split by $\sim2-3$~MHz. The finite detunings limits the two-qubit gate rates and fidelities, and can be improved by composite pulses and pulse shape engineering~\cite{Rong_Experimentl_2015, Vandersypen_NMR_2005, Dolde_high-fidelity_2014, Chen2016}. Finally, magnetic field orientation can be optimized to achieve higher optical cyclicities for both electron spin and nuclear spins~\cite{clear_optical-transition_2024} to realize high-fidelity single-shot readout of the qubit register. 

Although this work focused on a strongly coupled $^{29}$Si, the T center also interacts with a large bath of weakly coupled $^{29}$Si nuclei. The $S = \frac{1}{2}$ ground state of T center makes it challenging to control weakly coupled nuclear spins individually with dynamical decoupling (DD) gates.
However, recently developed dynamically decoupled radio frequency (DDRF) gates~\cite{beukers_control_2025} increase the control fidelity of nuclear spins with $S = \frac{1}{2}$. Leveraging the DDRF gates and hyperfine interactions with proximal nuclei (Fig. \ref{fig:SI_Azz_Axz_dist}), it might be possible to extend the qubit register size to ten qubits~\cite{bradley_ten-qubit_2019}. Our demonstration of a multi-qubit register combining $100$~ms memory coherence time and hyperfine interactions at MHz rates in a scalable, telecom-band photonics platform will find applications in quantum communication leveraging multi-qubit memory nodes \cite{Muralidharan2016} and efficient resource state generation for photonic measurement-based quantum computation~\cite{russo_generation_2019}.

\noindent{}\textbf{Acknowledgments.}
We thank Cameron Afradi and Kadircan Godeneli for experimental assistance, as well as Andrei Faraon for feedback on the manuscript. This work was primarily supported the U.S. Department of Energy, Office of Science, Basic Energy Sciences in Quantum Information Science under Award Number DE-SC0022289 for qubit synthesis and characterization, and first-principles modeling. L.K., Y.Z., and A.S acknowledge support from the NSF (QLCI program through grant number OMA-2016245, and  Award No. 2137645). We acknowledge additional support for cryogenic instrumentation  by the Office of Advanced Scientific Computing Research (ASCR), Office of Science, U.S. Department of Energy, under Contract No. DE-AC02-05CH11231 and Berkeley Lab FWP FP00013429. X.Z. acknowledges support from the Miller Institute for Basic Research in Science. The devices used in this work were fabricated at the Berkeley Marvell NanoLab and the Molecular Foundry at Berkeley Lab.

\bibliography{references}%

\appendix

\setcounter{figure}{0}
\renewcommand{\figurename}{Fig.}
\renewcommand{\thefigure}{S\arabic{figure}}
\renewcommand{\thetable}{S\arabic{table}}

\section{\label{SI: set up}Experimental setup}
\begin{figure*}
    \centering
    \includegraphics[width=\textwidth]{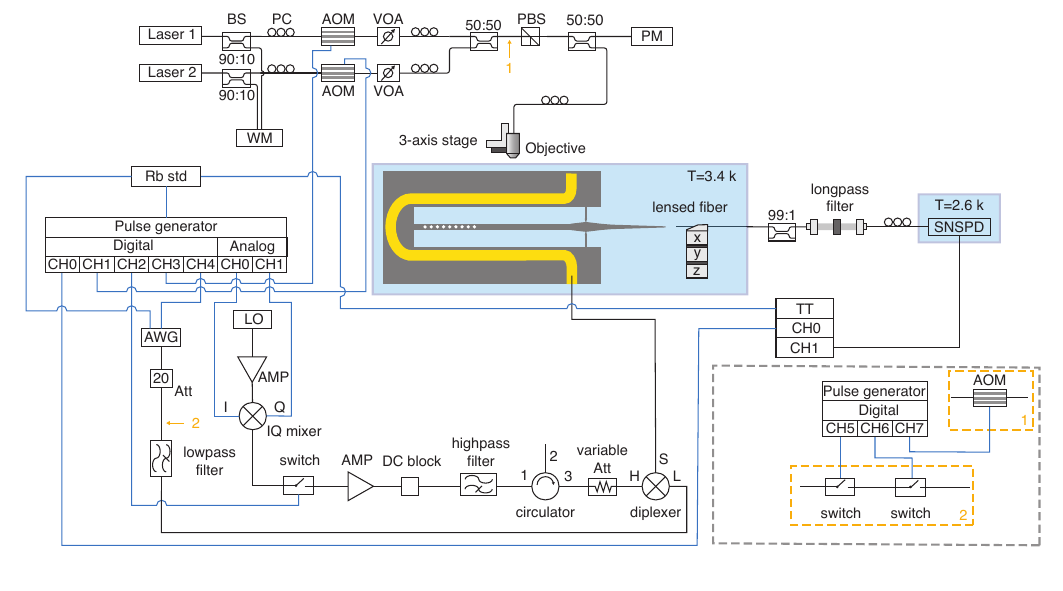}
    \vspace{-10mm}
    \caption{\textbf{Experimental setup.} Black lines represent optical fibers and coaxial cables. Blue lines represent coaxial cable connections for sequence control and synchronization. BS: Beam splitter. PC: Polarization controller. AOM: Acousto-optic modulator. VOA: Variable optical attenuator. PBS: Polarizing beam splitter. WM: wavemeter. PM: Power meter. SNSPD: Superconducting nanowire single photon detector. TT: Time tagger. Rb Std: Rubidium frequency standard. AWG: Arbitrary waveform generator. Att: Attenuator. LO: Local oscillator. AMP: amplifier. The dashed box shows extra setup for nuclear Hahn echo measurement.}
    \label{fig:setup}
\end{figure*}

The sample is mounted in a cryostat (Montana Instruments Cryostation s200) and cooled down to 3.4~K. We fix a Halbach array, assembled with three permanent magnets (K\&J Magnetics B888-2PA-N52 and B888-2PE-N52), below the sample to provide the external magnetic field. A lensed fiber (OZ Optics TSMJ-X-1550-9/125-0.25-7-3.5-19-2) is mounted on a 3-axis nanopositioner (Attocube ANPx101/LT and ANPz102/LT) for coupling to the photonic waveguide (efficiency $\sim 30\%$). Photons are detected with a superconducting nanowire single photon detector (Quantum Opus QO-NPD-1200-1600) with a quantum efficiency of $\sim 60\%$. Fig.~\ref{fig:setup} shows the experimental configuration. For free space excitation, the laser beam is focused by a microscope objective (Mitutoyo LCD Plan Apo NIR 50, $\mathrm{NA=0.42}$) outside the cryostat and sent through the top vacuum window of the cryostat. The objective is mounted on a 3-axis motorized micromanipulator (Sutter Instrument MP-285) for scanning.

To locate T centers, we spatially scan a free space continuous-wave laser at 635~nm (Thorlabs S1FC635) to excite T centers off-resonantly and collect fluorescence centered at 1325.4~nm (1.2~nm bandwidth) using a tunable bandpass filter (WL Photonics WLTF-BA-U-1310-100-SM-0.9/2.0-FC/APC-USB).
For photoluminescence excitation (PLE) measurements, we use two tunable O-band lasers (Santec TSL-570) for resonant excitation. Each laser is intensity modulated by a fiber-coupled acousto-optic modulator (AOM, Aerodiode 1310-AOM-2 for laser 1, IntraAction FCM-40.8E5C for laser 2). The laser power is measured using a power meter (ThorLabs PM100USB) with a fiber photodiode power sensor (ThorLabs S154C). The laser power is controlled independently with variable optical attenuators (Agiltron MSOA-02B1H1333). We combine the two laser beams with a 50:50 fiber beam splitter and use a fiber-coupled polarizing beam splitter to set identical polarization for the two lasers. A polarization controller is added before the objective to optimize the laser excitation efficiency based on T center signal. In the collection path, we use a free space longpass filter (ThorLabs FELH1350) to filter out the resonant excitation photons and collect fluorescence in phonon side band (PSB).

Microwave (MW) pulses are used to drive electron spin transitions. We first amplify the MW power from a local oscillator (Agilent Technologies E8257D) using a linear amplifier (Mini-Circuits ZX60-83LN-S+). The amplified LO signal is sent to the LO port of an IQ mixer (Marki microwave MMIQ-0218LXPC) and mixed with I and Q signals generated by the analog channels of the Swabian Instruments Pulse Streamer 8/2 for phase modulation and frequency upconversion. We add a MW switch (Analog Devices HMC547ALP3E) after the IQ mixer to attenuate LO leakage when IQ modulation is off. The upconverted MW signal is amplified with a linear amplifier (Mini-Circuits ZRON-8G+) and filtered by a DC block and a high pass filter (Mini-Circuits VHF-3100+). We use a circulator (DigiKey SFI4080A) after the filter to protect the amplifiers from reflected power. A digital attenuator (Vaunix LDA-908V) controls the power of the upconverted MW pulses. The MW signal is sent to the H port of a diplexer (Mini-Circuits ZDSS-3G4G-S+).

Radio-frequency (RF) pulses are used to drive nuclear spin transitions. We use an arbitrary waveform generator (AWG: Siglent SDG6022X) to directly synthesize the RF signal and then attenuate it with a 20 dB attenuator to achieve higher amplitude resolution. After going through a lowpass filter (Mini-Circuits VLF-80+), the RF signal is sent to the L port of the diplexer. The combined MW and RF signals are sent to a custom printed circuit board (PCB) which is wire bonded the niobium (Nb) transmission line on the chip inside the cryostat. 

When measuring nuclear spin Hahn echo coherence times, we add an extra AOM (Aerodiode 1310-AOM-2) before the polarizing beam splitter to further reduce laser leakage. We also change the RF line setup to the configuration shown in the dashed box in Fig.~\ref{fig:setup}. The AWG is used as local oscillator and two MW switches (Mini-Circuits ZASWA-2-50DRA+) are cascaded to pulse RF signals to accommodate memory limitations on the AWG. We keep other configurations the same.

Pulses are generated using a Swabian Instruments Pulse Streamer 8/2. The photon clicks from SNSPD are time-tagged with a Swabian Instruments Time Tagger Ultra. Pulse streamer, time tagger and AWG are all synchronized to a reference Rubidium clock (Stanford Research Systems Rubidium Frequency Standard).

\section{\label{SI: fabrication}Device fabrication}

\begin{figure*}
    \centering
    \includegraphics[]{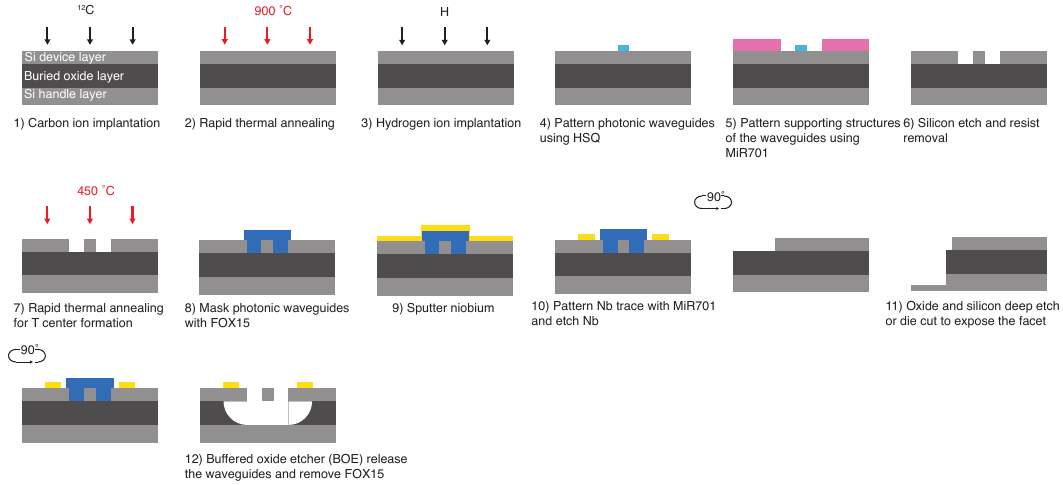}
    \caption{\textbf{Device fabrication.} 1-3, 7) Ion implantation and rapid thermal annealing for T center generation. 4-6) Electron-beam lithography and optical lithography steps for defining photonic waveguides. 8-10) Niobium sputtering and optical lithography for metal transmission line. We use electron-beam lithography to pattern $\mathrm{SiO_2}$ masks (FOX15) to protect photonic waveguides during metal etching. 11) Side view of the photonic waveguide, the tapered waveguide end of the chip is exposed for fiber coupling. 12) Wet release of the device with buffered oxide etcher (BOE). FOX15 masks are also removed with BOE.}
    \label{fig:fab}
\end{figure*}
We fabricate our device on a 1~cm $\times$ 1~cm chip diced from a 200 mm silicon-on-insulator (SOI) wafer (SEH America). The 220~nm thick, float-zone grown device layer has a high resistivity ($\ge 3000~\mathrm{\Omega~cm}$) and (100) orientation. To create T centers, we first implant $^{12}$C ions at 36 keV (average depth $110$~nm) with a fluence of $1\times10^{12}$~cm$^{-2}$ and a $7^{\circ}$ tilt. The implantation is followed by rapid thermal annealing in $\mathrm{N_2}$ atmosphere at 900~$^{\circ}$C for 20 seconds to repair lattice damage from implantation. We then implant H at 9~keV (average depth $110$~nm) with a fluence of $7\times10^{12}$~cm$^{-2}$, and anneal the device at 450~$^{\circ}$C in $\mathrm{N_2}$ atmosphere for 3 min to form T centers. Details for device fabrication and T center creation processes are shown in Fig.~\ref{fig:fab}. The single mode photonic waveguide is along the $\langle 110 \rangle$ direction and is terminated with a Bragg mirror at the end, following the design presented in Ref.~\cite{komza_multiplexed_2025}. After device fabrication, we observe an average of one T center per 60~$\mu$m of the waveguide ($400~$nm width), which corresponds to a T center concentration of 0.004 ppb in silicon. For early stage measurements, we have studied a chip with a higher carbon implantation density ($1\times10^{13}$~cm$^{-2}$). With the same hydrogen implantation density, we measure a density of one T center per 0.2~$\mu$m of waveguide ($400~$nm width) in that chip, which corresponds to a T center concentration of $0.7$~ppb in silicon.

\section{\label{SI: silicon hyperfine}First principles calculations of hyperfine  interaction with silicon}
\begin{figure*}
    \centering
    \includegraphics[]{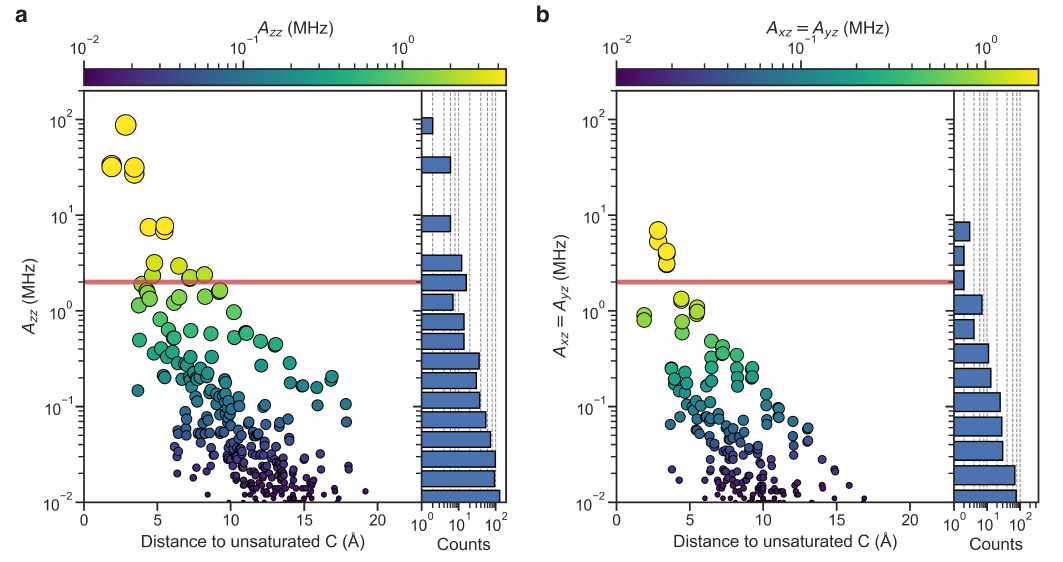}
    \caption{\textbf{Distribution of hyperfine coupling strength between $^{29}$Si and the T center electron spin}. \textbf{a,} Distribution of diagonal $A_{zz}$ of $^{29}$Si with respect to the distance to the unsaturated carbon of T center. The color bar and marker size stand for the values of $A_{zz}$. The horizontal line stands for $A_{zz}=2$~MHz. Every lattice site is populated with $^{29}$Si. \textbf{b,} Distribution of diagonal $A_{xz}=A_{yz}$ of $^{29}$Si with respect to the distance to the unsaturated carbon of T center. The color bar and marker size stand for the values of $A_{xz}=A_{yz}$. The horizontal line stands for $A_{xz}/A_{yz}=2$~MHz.}
    \label{fig:SI_Azz_Axz_dist}
\end{figure*}
The hyperfine simulations are performed using VASP~\cite{G.Kresse-PRB96,G.Kresse-CMS96} and the projector-augmented wave (PAW) method~\cite{P.E.Blochl-PRB94}. T center defect structure is simulated using a 1002-atom supercell. In our setup, the C-C axis is aligned along the $Z$ axis, with the C-C-H defect lying in the [110] defect plane. Structural optimizations are performed with fixed supercell volumes until the ionic forces are smaller than 0.01~eV/\r{A}. Spin-polarized Heyd-Scuseria-Ernzerhoff (HSE)~\cite{Heyd2003} with 25\% exact exchange is employed to simulate its doublet ground state. The hyperfine coupling constants are evaluated using the implementation in VASP. The hyperfine tensor $A_{ij}^I$ of nucleus $I$ comprises the isotropic Fermi-contact term

\begin{equation}
{A_{\mathrm{iso}}^{I}}=\frac{2}{3} \frac{\mu_{0} \gamma_{e} \gamma_{I}}{S} \delta_{i j} \int \delta_{T}(\mathbf{r}) \rho_{s}\left(\mathbf{r}+\mathbf{R}_{I}\right) d \mathbf{r},
\end{equation}
and the anisotropic spin dipolar term
\begin{equation}
{A_{\mathrm{ani}}^{I}}=\frac{\mu_{0}}{4 \pi} \frac{\gamma_{e} \gamma_{I}}{S} \int \frac{\rho_{s}\left(\mathbf{r}+\mathbf{R}_{I}\right)}{r^{3}} \frac{3 r_{i} r_{j}-\delta_{i j} r^{2}}{r^{2}} d \mathbf{r},
\end{equation}
where $\mu_{0}$ is the permeability of vacuum, $\gamma_{e}$ and $\gamma_{I}$ are the gyromagnetic ratios of the electron and nuclei. $\delta_{T}(\mathbf{r})$ is a smeared out $\delta$ function in the relativistic case~\cite{blochl2000}. $\rho_{s}$ is the spin density of spin state $S$ at coordinates $\mathbf{r}$ with respect to the position of the nucleus $\mathbf{R}_{I}$.

The spin quantization axis in simulations is defined along the carbon–carbon bond of the T center, corresponding to the $z$-axis in our setup. We populate all the silicon lattice sites with $^{29}$Si atoms in a 1000-atom supercell. The distribution of hyperfine couplings is plotted in terms of the diagonal ($A_{zz}$) and off-diagonal ($A_{xz}=A_{yz}$) terms in Fig.~\ref{fig:SI_Azz_Axz_dist}\textbf{a} and Fig.~\ref{fig:SI_Azz_Axz_dist}\textbf{b}, respectively. The horizontal line indicates hyperfine values larger than 2 MHz. This statistical analysis can guide the estimation of $^{29}$Si distribution. For instance, it shows that 28 lattice sites within approximately 10~\AA~have $A_{zz}>2$ MHz. Given that the simulation contains 1000 atoms, we estimate approximately 1.3 $^{29}$Si atoms have $A_{zz}>2$ MHz within approximately 10\AA~ for natural silicon (28 sites $\times$ 4.66\% natural abundance of $^{29}$Si).

\begin{figure}
    \centering
    \includegraphics[width=1\columnwidth]{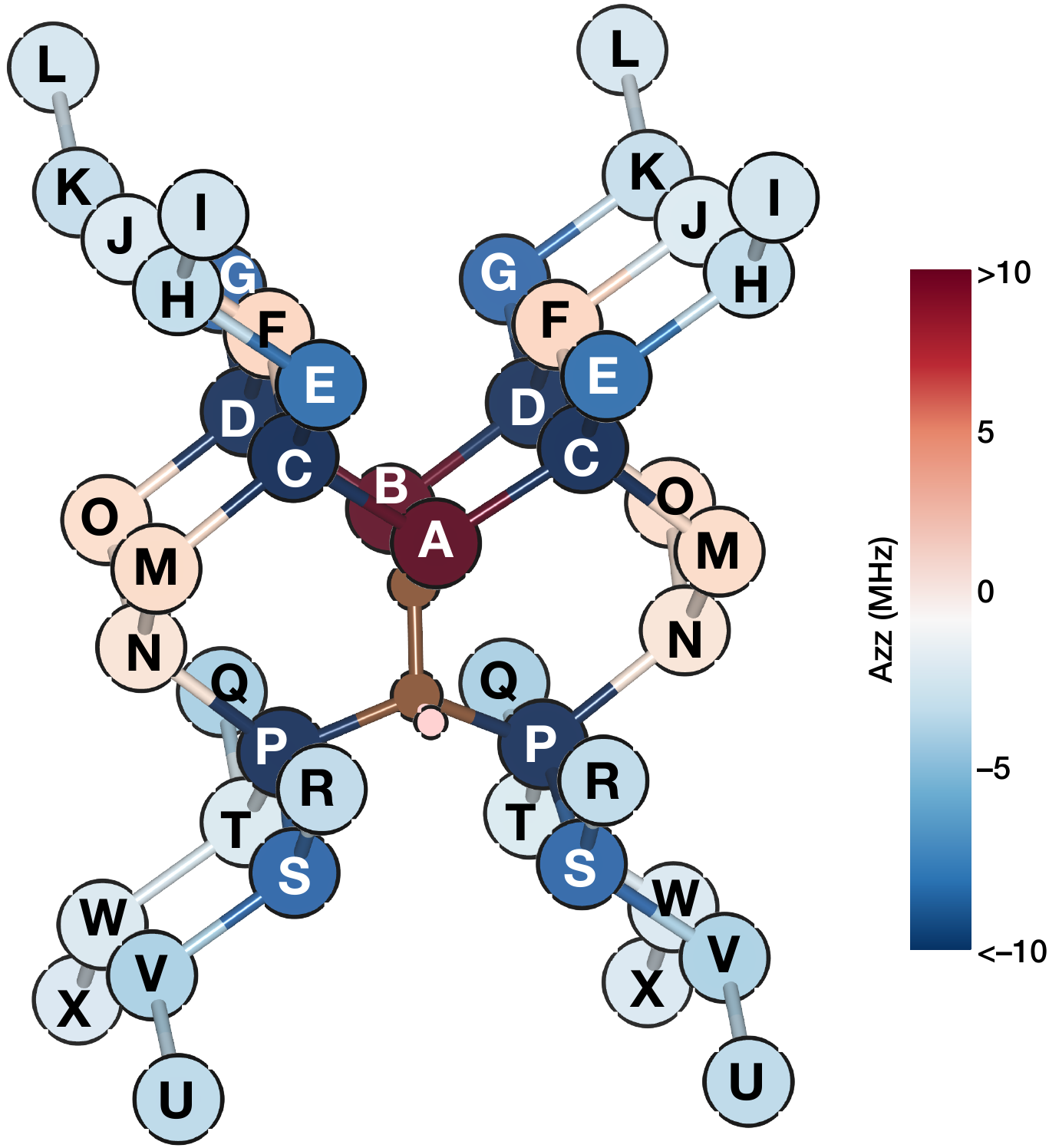}
    \caption{Defect structure of a T center where only the atoms with $|A_{zz}|>1$~MHz are presented. The diagonal $A_{zz}$ values are visualized by the color bar. The defect structure shows a $C_{1h}$ symmetry along the C–C–H plane. This is also reflected on the labeling of the atoms.}
    \label{fig:SI_Azz_struct}
\end{figure}

\begin{table}[h]
    \caption{\label{tab:atoms}Values for diagonal $A_{zz}$ (MHz) and off-diagonal $A_{xz}=A_{yz}$ (MHz) for atoms A-X for that have $|A_{zz}|>1$~MHz. In $A_{xz}=A_{yz}$ column, the values separated by / stands for the atoms with the same labeling on the left and right of the defect [110] plane, respectively.}
    \begin{ruledtabular}
    \begin{tabular}{ccc}
Atom & $A_{zz}$ (MHz) & $A_{xz}=A_{yz}$ (MHz) \\ \hline
A    & 31.9    & 0.8             \\
B    & 33.31   & -0.91           \\
C    & -27.22  & 3.96 / -3.06    \\
D    & -31.58  & 3.12 / -4.16    \\
E    & -6.89   & -0.93 / -1.05   \\
F    & 1.9     & 0.16 / -0.17    \\
G    & -7.7    & 1.09 / 0.98     \\
H    & -2.19   & 0.42 / -0.35    \\
I    & -1.59   & -0.25 / -0.2    \\
J    & -1.22   & 0.17 / -0.16    \\
K    & -2.23   & 0.36 / -0.42    \\
L    & -1.63   & 0.2 / 0.25      \\
M    & 1.62    & -0.02 / -0.14   \\
N    & 1.14    & -0.25 / 0.25    \\
O    & 1.54    & 0.08 / -0.00    \\
P    & -87.61  & -6.93 / 5.31    \\
Q    & -3.18   & -0.17 / -0.10   \\
R    & -2.33   & 0.14 / 0.23     \\
S    & -7.47   & 1.34 / 1.28     \\
T    & -1.34   & -0.77 / -0.59   \\
U    & -2.38   & 0.35 / 0.25     \\
V    & -2.94   & -0.48 / 0.32    \\
W    & -1.38   & -0.18 / 0.26    \\
X    & -1.41   & -0.16 / -0.2    \\
    \end{tabular}
    \end{ruledtabular}
    \label{table:A_tensor_table}
\end{table}

A defect structure of the T center is shown in Fig.~\ref{fig:SI_Azz_struct}; only atoms with $A_{zz}$ larger than 1~MHz are shown for simplicity. The $A_{zz}$ values of $^{29}$Si atoms are visualized using a color code, with detailed numerical values summarized in Table~\ref{table:A_tensor_table}. The distribution of $A_{zz}$ along the quantization axis reflects the defect’s $C_{1h}$ symmetry, as the values are mirrored across the C–C–H defect plane. The complete hyperfine tensor dataset simulated in this work is publicly available on Zenodo \cite{zenodo_repo}.

\section{\label{SI:otherT} Observation of strongly coupled nuclear spins on other T centers}

\begin{figure}
    \centering
    \includegraphics[]{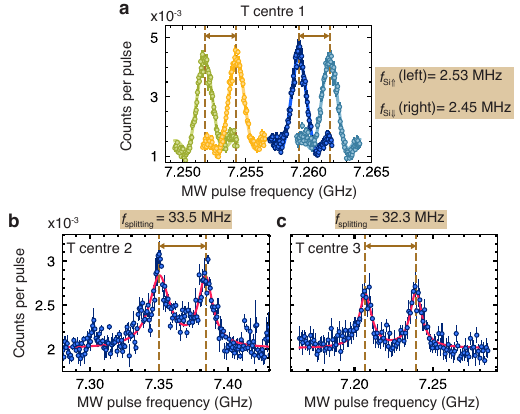}
    \caption{\textbf{T centers with more than one strongly coupled nuclear spin.} The pulsed-ODMR spectra of T centers are taken under the same permanent magnetic field. The splitting of the electron spin transition is marked on each figure. \textbf{a,} Electron spin resonance of T center 1 (T center in the main text), plotted here for reference. \textbf{b,} A T centerf (T center 2) in a chip with high carbon implantation density. \textbf{c,} A T center (T center 3) in the same chip (low carbon implantation density) as T center 1. }
    \label{fig:SI_otherT}
\end{figure}

We  investigated three T centers under the same magnetic field configuration (Fig.~\ref{fig:SI_otherT}): T centers 1 and 3 are in a low carbon implantation density chip, and T center 2 is in a high carbon implantation density chip. With only electron spin initialization of T centers 2 and 3, their pulsed-ODMR spectra contain two peaks that are separated by around $33~$MHz for both T centers (Fig.~\ref{fig:SI_otherT}\textbf{b, c}). The magnitude of the splitting is much greater than the expected hydrogen hyperfine interaction induced splitting in electron spin resonance, estimated to be several MHz based on the alignment between the T center and the external magnetic field~\cite{bergeron_silicon-integrated_2020, clear_optical-transition_2024}. As a reference, we measure a hydrogen hyperfine interaction induced splitting of 2.5 MHz in the T center in main text (Fig.~\ref{fig:SI_otherT}\textbf{a}). Therefore, we conclude that the $33~$MHz splitting in T centers 2 and 3 is likely a result of T center electron spin interacting with another nuclear spin that has a large hyperfine coupling. The high possibility to observe a strongly coupled nuclear spin in addition to hydrogen agrees with the DFT calculated $^{29}$Si hyperfine tensor in Appendix \ref{SI: silicon hyperfine}. Based on  DFT calculations, the possibility of finding a T center with more than one strongly coupled nuclear spin is close to unity, considering 46 lattice sites show hyperfine coupling $> 1~$MHz (Table~\ref{table:A_tensor_table}) and the 4.67\% isotope ratio of $^{29}$Si in natural silicon.

\section{\label{SI:search ESR}Identification of nuclear spin dependent electron spin transitions}

\begin{figure}
    \centering
    \includegraphics[]{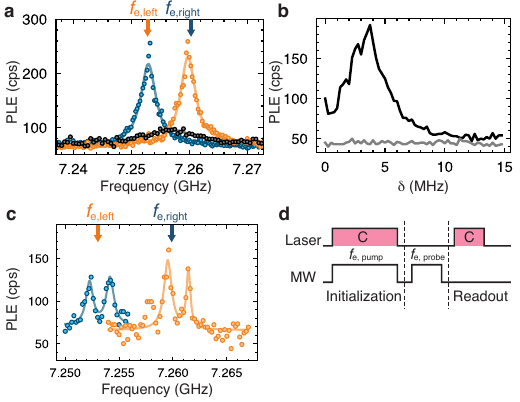}
    \caption{\textbf{Identification of electron spin resonances.} \textbf{a,} Two-tone continuous-wave (CW) ODMR measurements. The arrows indicate the pump MW frequency, while the traces sharing the same color show the ESR response when sweeping the probe MW frequency. The two ESR peaks are separated by 6.9~MHz and the linewidth of each peak is 3.1~MHz. Single-tone CW-ODMR is shown as a reference (black). \textbf{b,} Two-tone CW-ODMR showing a resonance at $\delta = 3.75$~MHz (black). The detuning of the two MW tones $2\delta$ are swept while keeping the center frequency $f_0=7.257$~GHz fixed. The dark count is shown as a reference (gray). \textbf{c,} Pulsed-ODMR. The arrows indicate the MW frequencies used in initialization ($f_{\mathrm{e,~pump}}$). The traces sharing the same color show the pulsed-ODMR response. Each ESR peak observed in \textbf{a} is further split to two peaks. The splitting between two small peaks within the same trace is 2.5~MHz. \textbf{d,} Pulse sequence for \textbf{c}.}
    \label{fig:SI_search_ESR}
\end{figure}

For T centers, the combined effect of nuclear Zeeman splitting and nuclear-electron hyperfine interactions results in nuclear spin dependent electron spin resonances (ESRs). If the nuclear spins are not initialized, we observe a broad and small peak at $f_0=7.257$~GHz in CW-ODMR~(Fig.~\ref{fig:SI_search_ESR}\textbf{a}, black), consistent with having a single MW tone off-resonantly driving multiple ESRs. Therefore, we use two-tone ODMR to address different nuclear spin states. First, we sweep the detuning ($2 \delta$) of the two MW tones while keeping their average frequency fixed at $f_0$. We observe a two-tone resonance at  $\delta=3.75$ MHz ~(Fig.~\ref{fig:SI_search_ESR}\textbf{b}). We then move on to pump-probe ODMR. With one MW tone fixed at $f_0+\delta$, scanning the second MW tone shows a resonance at  $(f_{\mathrm{e, left}}\approx f_0-\delta$ ). By optimizing the two MW frequencies iteratively based on the resonance contrast, we identify two ESRs in cw-ODMR (Fig.~\ref{fig:SI_search_ESR}\textbf{a}). To probe finer structures in ESR, we reduce the MW driving power. With the pump-probe sequence (Fig.~\ref{fig:SI_search_ESR}\textbf{d}), each ESR peak further splits into two peaks (Fig.~\ref{fig:SI_search_ESR}\textbf{c}), suggesting the T center under study is coupled to a second nuclear spin in addition to its intrinsic hydrogen nuclear spin. With a full mapping of the electron spin transitions, we develop the pulse sequence shown in Fig.~\ref{fig:fig1}\textbf{f} for initialization and optimize the MW driving frequencies based on the resonance contrast.

\section{\label{SI:freq_jump}Meissner effect induced electron spin resonance shifts}
\begin{figure}
    \centering
    \includegraphics[]{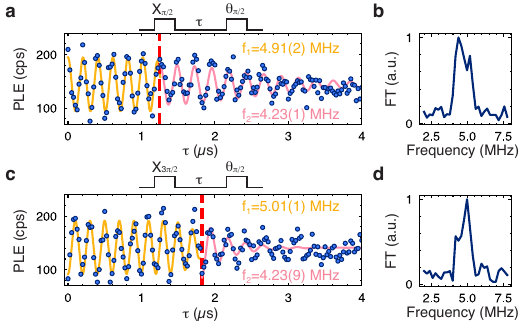}
    \caption{\textbf{ESR shifts due to Meissner effect.} \textbf{a,} Ramsey fringes of the electron spin. A $\Delta=5$ MHz virtual detuning  is added to the second $\pi/2$ pulse phase  $\theta=\Delta \tau$. We fit an oscillation frequency at 4.91(2)~MHz for $\tau<1.25~\mathrm{\mu s}$ and at 4.23(1)~MHz for $\tau>1.25~\mathrm{\mu s}$. \textbf{b,}  Fourier Transform of the raw data in \textbf{a} showing two frequency components of the oscillation. \textbf{c,} Ramsey fringe using a $3\pi/2$ pulse as the first pulse while applying the same virtual rotation on second $\pi/2$ pulse as discussed in \textbf{a}. The oscillation frequency of the signal changes at $\tau=1.825~\mathrm{\mu s}$. \textbf{d,} DFT of the signal trace in \textbf{c} showing two frequency components in the oscillation.}
    \label{fig:Meissner}
\end{figure}
During early stages of our pulsed-ODMR experiments, we observed transient frequency shifts in electron spin resonance (ESR) that depended on the microwave pulse power and duration. To understand such shifts, we use Ramsey sequences to measure their dynamics. We add a 5 MHz virtual detuning to the second $\pi/2$ pulse of the Ramsey sequence~(Fig.~\ref{fig:Meissner}\textbf{a}). By analyzing the oscillation frequency in the signal and comparing it against 5~MHz, we can extract the precise ESR. Surprisingly, we notice two frequency components in the Ramsey signal oscillation (Fig.~\ref{fig:Meissner}\textbf{b}) and find that the frequency jump occurs around $1.25~\mathrm{\mu s}$ after the first $\pi/2$ pulse. 

We hypothesize that this frequency jump is due to the Meissner effect in superconductors. In our measurements, we use a superconducting transmission line to minimize the heating effects of MW and RF control pulses. However, if the metal film undergoes superconducting to normal phase transition, the change in magnetic screening  by the metal trace would induce a small change in the local magnetic field seen by the T center. In Ramsey measurements, if the AC current for the initial $\pi/2$ pulse is above the critical current of our metal film, the film turns normal. During the delay time ($\tau$), the film thermalizes with the cryostat and returns to superconducting phase after a delay. When the phase transition occurs, the Meissner effect leads to a part per ten thousand  change (shift:  $0.7 $ MHz,  carrier frequency: $7.3$ GHz) of the local magnetic field and the T center ESR, which appears as a frequency jump in Ramsey signal~(Fig.~\ref{fig:Meissner}\textbf{a}). We verify this hypothesis by changing the initial $\pi/2$ pulse to a $3\pi/2$ pulse for more heating. We observe that the frequency jump occurs later with the longer pulse, which is well explained by the longer time needed for the metal film to cool down to below critical temperature after more heating.

To maintain the phase of Nb film and avoid frequency jumps in our measurements, we carefully choose our experimental parameters such that the metal film is either always superconducting or always normal in any given measurement. During the measurement in Fig.~\ref{fig:fig2}\textbf{b}, we send in a low frequency RF signal (30 MHz) that is detuned from the nuclear spin resonances to keep the metal film normal. Similarly, we also keep the metal film normal for measurements shown in Fig.~\ref{fig:fig2}\textbf{c}. 

\section{\label{SI:pulse fidelity} Pulse fidelity for electron  and nuclear spins} 
\begin{figure*}
    \centering
    \includegraphics[]{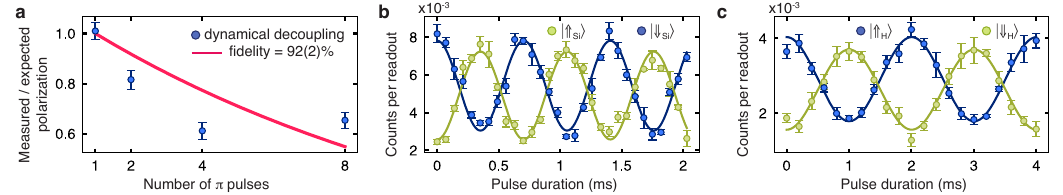}
    \caption{\textbf{Electron spin and nuclear spin pulse fidelity. a,} Electron spin $\pi$ pulse fidelity estimated from dynamical decoupling measurements. The average polarization of the first ten data points of each dynamical decoupling trace are plotted against the number of $\pi$ pulses for fitting. \textbf{b,} Silicon nuclear spin Rabi oscillation. We estimate a fidelity of 89(5)\% for Si $\pi/2$ pulses. \textbf{c,} Hydrogen nuclear spin Rabi oscillation. We estimate a fidelity of 88(4)\% for H $\pi$ pulses.}
    \label{fig:pulse_fidelity}
\end{figure*}

We estimate an electron spin $\pi$ pulse fidelity of 92(2)\% (Fig.~\ref{fig:pulse_fidelity}\textbf{a}) using data from electron spin dynamical decoupling measurements (Fig.~\ref{fig:fig2}\textbf{c}) and assuming uncorrelated errors. We use the Rabi oscillation of silicon and nuclear spins shown in Fig.~\ref{fig:pulse_fidelity}\textbf{b,~c} to construct the pulses for Bell state generation and estimate nuclear pulse errors. 

\section{\label{SI:cross_transitions}Nuclear spin non-conserving electron spin transitions}
\begin{figure}
    \centering
    \includegraphics[]{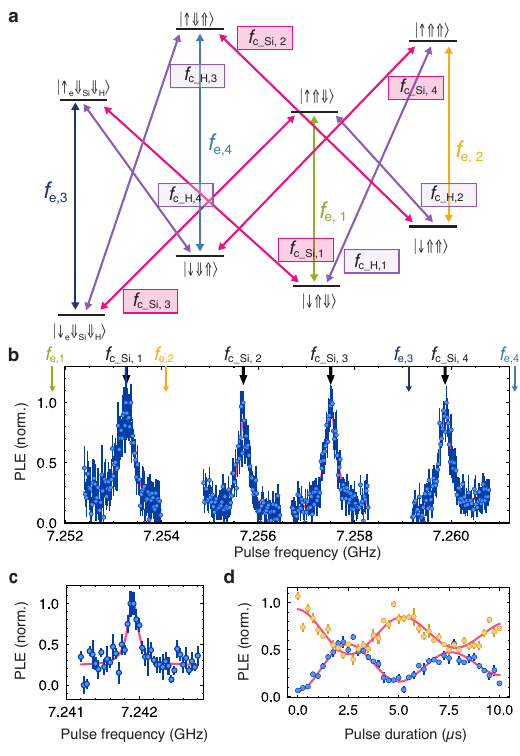}
    \caption{\textbf{Nuclear spin non-conserving electron spin transitions.} \textbf{a, }A zoomed-in view of ground state level structure of the T center. \textbf{b, }Pulsed-ODMR spectra of the silicon nuclear spin flipping ESRs. The arrows indicate nuclear spin conserving ESRs. \textbf{c, }Pulsed-ODMR spectrum of a hydrogen nuclear spin flipping ESR ($f_{\mathrm{c\_H}, 2}$). \textbf{d, }Rabi oscillation of the electron spin with MW driving a silicon nuclear spin flipping ESR ($f_{\mathrm{c\_Si}, 3}$). The system is initialized in $\mathrm{\lvert\downarrow_e\Downarrow_{Si}\Downarrow_{H}\rangle}$, the population in $\mathrm{\lvert\downarrow_e\rangle}$ is shown in yellow dots, while the population in $\mathrm{\lvert\uparrow_e\rangle}$ is shown in blue dots.} 
    \label{fig:SI_cross_transitions}
\end{figure}

Due to the off-diagonal components in hyperfine interaction, the nuclear spin non-conserving electron spin transitions are weakly allowed. We measure those transitions and realize that they are close to nuclear spin conserving ESRs. In Fig.~\ref{fig:SI_cross_transitions}\textbf{b}, we show silicon nuclear spin flipping ESRs ($f_{\mathrm{c\_Si}, 1-4}$). In Fig~\ref{fig:SI_cross_transitions}\textbf{c}, we show hydrogen nuclear spin flipping ESR ($f_{\mathrm{c\_H}, 2}$). Driving nuclear spin flipping electron spin transitions also allow coherent control. An example of Rabi oscillation by driving $f_{\mathrm{c\_Si}, 3}$ is shown in Fig.~\ref{fig:SI_cross_transitions}\textbf{d}.


\section{\label{SI:Up_e NMR}NMR in $\lvert\uparrow_e\rangle$ manifold}

\begin{figure}
    \centering
    \includegraphics[]{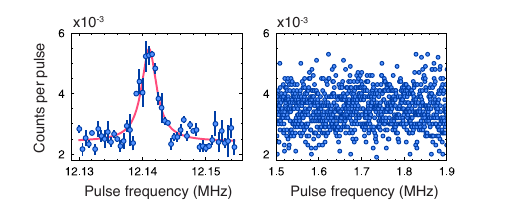}
    \caption{\textbf{Nuclear spin resonances in $\lvert\uparrow_e\rangle$ manifold.} \textbf{a,} Pulsed-ODMR for hydrogen nuclear spin resonance. \textbf{b,} Pulsed-ODMR around the predicted region for silicon nuclear spin resonance. No resonance is observed within our measurement sensitivity.}
    \label{fig:up_e_H}
\end{figure}

In the main text, we show the mapping of NMR transitions in $\lvert\downarrow_e\rangle$ manifold. We also investigate NMRs in $\lvert\uparrow_e\rangle$ manifold. The T center system is initialized in $\lvert\uparrow_{\mathrm{e}} \Downarrow_{\mathrm{Si}} \Downarrow_{\mathrm{H}} \rangle$ with optical pumping at transition B and MW pumping at $f_{e: 1,2,4}$. After initialization, we measure the transition resonance for hydrogen nuclear spin at 12.141~MHz~(Fig.~\ref{fig:up_e_H}\textbf{a}). Using the measured frequencies of nuclear spin conserving ESRs and nuclear spin transition resonances, we predict silicon NMR in $\lvert\uparrow_\mathrm{e}\rangle$ manifold at around 1.65~MHz. However, we are unable to detect silicon NMR peak in pulsed-ODMR centered around 1.65~MHz when initializing the T center system in the $\lvert\uparrow_{\mathrm{e}} \Uparrow_{\mathrm{Si}}  \Downarrow_{\mathrm{H}} \rangle$ state~(Fig.~\ref{fig:up_e_H}\textbf{b}). 

The lack of a silicon NMR in the $\lvert\uparrow_\mathrm{e}\rangle$ manifold requires further investigation. It is possible that the quantization axis of the silicon nuclear spin deviates significantly from the external static magnetic direction ($\vec{\text{B}}_{\text{DC}}$) due to the transverse hyperfine interaction. Since the AC driving field direction ($\vec{\text{B}}_{\text{AC}}$) is roughly perpendicular to $\vec{\text{B}}_{\text{DC}}$, a large deviation of the nuclear spin quantization axis from $\vec{\text{B}}_{\text{DC}}$ will lead to inefficient driving and challenges in observing Si NMR peak in the $\lvert\uparrow_{\mathrm{e}}\rangle$ manifold.


\section{\label{SI:Tomography}Phase reversal tomography}
\begin{figure}
    \centering
    \includegraphics[]{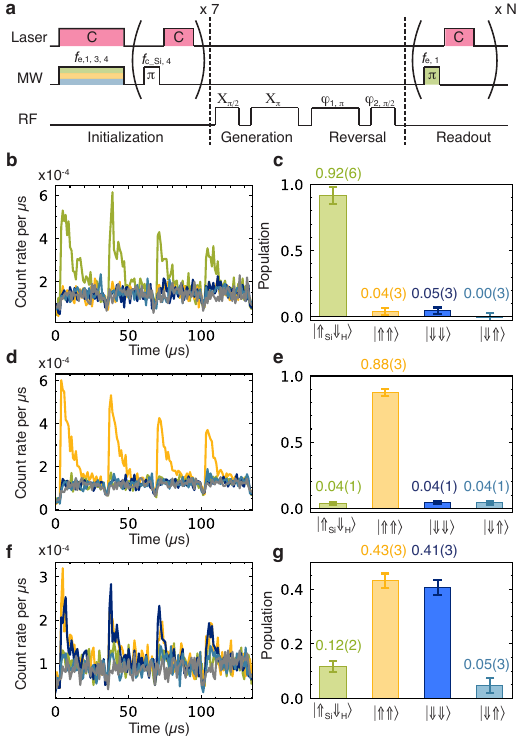}
    \caption{\textbf{State initialization, entangled state generation and tomography.} \textbf{a, }Pulse sequence for Bell state generation and phase reversal tomography. The sequence initializes the system in the $\lvert\downarrow\Downarrow\Downarrow\rangle$ state. Compared to Fig.~\ref{fig:fig1}\textbf{f}, we add an additional step in initialization: we apply nuclear spin-flipping electron spin $\pi$ pulse that maps the register to $\mathrm{\lvert\uparrow\Downarrow\Downarrow\rangle}$, and followed by a laser pulse that pump the electron spin to $\mathrm{\lvert\downarrow\rangle}$ state. \textbf{b, }Examples of population readout traces when the system is initialized in $\lvert\downarrow\Uparrow\Downarrow\rangle$. The readout is repeated for each state. The background readout signal (gray) is shown as a reference. \textbf{c,} Calculated population of each state after background subtraction based on measurement results in \textbf{b}. \textbf{d,~e,} Population measurement when the system is initialized in $\lvert\downarrow\Uparrow\Uparrow\rangle$ state. \textbf{f,~g, }Population measurements after Bell state, $\lvert\Phi^-\rangle$, generation. The equal amount of population in $\lvert\Uparrow\Uparrow\rangle$ and $\lvert\Downarrow\Downarrow\rangle$ demonstrates correlation between the two nuclear spins.}
    \label{fig:SI_tomography}
\end{figure}

Our system of strongly coupled electron and nuclear spins naturally provides various CNOT gates to generate entangled states between nuclear spins. However, the controlled nature of the gates (i.e. no true single qubit gate available) presents significant challenges in performing full state tomography of the nuclear spin register. Therefore, we use phase reversal tomography, a partial tomography technique, to benchmark our entanglement fidelity~\cite{mehring_entanglement_2003, stemp_tomography_2024}.

Realizing high fidelity entanglement requires high fidelity state preparation. In addition to the initialization sequence shown in Fig.~\ref{fig:fig1}\textbf{f}, we include additional initialization sequences that map the nuclear spin to the target state using nuclear spin flipping electron spin transitions (Fig.~\ref{fig:SI_tomography}\textbf{a}). To quantify the initialization, we measure the population for each state ($\langle \sigma_z\rangle$) by counting the photons collected in the first two readout pulses ($N=2$). A reference readout sequence without MW $\pi$ pulses is used to calibrate the background counts of the measurement ($\mathrm{PL_{bg}}$). We then subtract $\mathrm{PL_{bg}}$ from the PL counts of each state and calculate the population of each state ($\lvert\alpha\beta\rangle$) by
\begin{equation}
    p_{\lvert\alpha\beta\rangle} = \frac{\mathrm{PL}_{\lvert\alpha\beta\rangle}}{\mathrm{PL}_{\lvert\Uparrow\Uparrow\rangle}+\mathrm{PL}_{\lvert\Uparrow\Downarrow\rangle}+\mathrm{PL}_{\lvert\Downarrow\Uparrow\rangle}+\mathrm{PL}_{\lvert\Downarrow\Downarrow\rangle}}
\end{equation}

For state tomography, the Bell state can be represented using a generic density matrix ($\rho_{exp}$):
\begin{equation}
    \begin{pNiceMatrix}[first-row, first-col]
            & \lvert\Downarrow\Downarrow\rangle & \lvert\Downarrow\Uparrow\rangle & \lvert\Uparrow\Downarrow\rangle & \lvert\Uparrow\Uparrow\rangle \\
        \lvert\Downarrow\Downarrow\rangle  & p_1 & a  & b  & c \\
        \lvert\Downarrow\Uparrow\rangle  & a^{*}  & p_2 & d  & e \\
        \lvert\Uparrow\Downarrow\rangle  & b^{*} & d^{*}  & p_3  & f \\
        \lvert\Uparrow\Uparrow\rangle  & c^{*} & e^{*}  & f^{*}  & p_4 \\
    \end{pNiceMatrix}
\end{equation}
An ideal Bell state only has four non-zero elements in its density matrix ($\rho_{\text{ideal}}$). For example, $\rho_{\text{ideal}}$ of $\lvert\Phi^-\rangle$ has $p_1 = p_4 = 0.5$ (population), $c = c^{*}=-0.5$ (coherence). For diagonal elements ($p_{1-4}$) of $\rho_{\text{exp}}$, we perform direct population measurements as discussed in Fig.~\ref{fig:SI_tomography}\textbf{d, e}. To measure the non-zero off-diagonal term $c = a+b\cdot i$ for $\lvert\Phi^-\rangle$, we use the phase reversal tomography where we measure the expectation value of silicon nuclear spin, $\langle \sigma_z\rangle_{\mathrm{Si}}$, as a function of phases in the reversal pulses where
\begin{equation}
    \langle \sigma_z\rangle_{\mathrm{Si}}=(p_3-p_2)-2 a \cdot \cos(\varphi_1+\varphi_2) + 2b\cdot\sin(\varphi_1+\varphi_2)
\end{equation}
In our measurement, we set $\varphi_1 = \theta$, $\varphi_2 = 3\theta$, and perform population measurement for each $\theta$. We define \mbox{$\langle\Downarrow\lvert\sigma_z\rvert\Downarrow\rangle = -1$} and  $\langle\Uparrow\lvert\sigma_z\rvert\Uparrow\rangle = 1$, therefore the expectation value of silicon qubit is
\begin{equation}
    \mathrm{\langle \sigma_z\rangle_{\mathrm{Si}}= -(PL_{\lvert\Downarrow\Uparrow\rangle}+PL_{\lvert\Downarrow\Downarrow\rangle})+(PL_{\lvert\Uparrow\Uparrow\rangle}+PL_{\lvert\Uparrow\Downarrow\rangle})}
\end{equation}
Using our measured data shown in Fig.~\ref{fig:fig4}\textbf{b}, we extract $c = -0.35(2) - i\cdot0.03(2) $ when preparing the nuclear spins to $\lvert\Phi^-\rangle$. We note that we cannot reconstruct the full density matrix using phase reversal tomography. However, the extracted elements related to population and coherence are sufficient for determining the fidelity
\begin{equation}
    F = \mathrm{Tr}(\rho_{\text{exp}}\times\rho_{\text{ideal}})
\end{equation}
For $\lvert\Phi^-\rangle$, the expression simplifies to $F = 0.5(p_1+p_4)-a$ and we estimate the fidelity to be $0.771\pm 0.028$. The measured fidelity can be accounted for by the combination of state initialization fidelity (Fig.~\ref{fig:SI_tomography}\textbf{e}) and pulse fidelity for entanglement generation (Fig.~\ref{fig:pulse_fidelity}\textbf{b,~c}) that results in an expected fidelity of 0.69(6).  

\section{\label{SI:Bell_coherence}Bell state coherence}
In this section, we estimate the expected Bell state coherence time using the coherence times of the two nuclear spins. 
We assume that both nuclear spins are subject to some magnetic noise that leads to phase accumulation of $\varphi_{\textrm{Si}}(t)$ for the Si nuclear spin and $\varphi_{\textrm{H}}(t)$ for H nuclear spin during Ramsey measurements. The entangled states experience the combined phase accumulation after a free evolution time t:

\begin{align}
    \lvert \Phi^\pm \rangle &: \frac{1}{\sqrt2}(\lvert 00 \rangle \pm e^{i(\varphi_{\textrm{Si}}(t) + \varphi_{\textrm{H}}(t) )} \lvert 11 \rangle )\\ \lvert \Psi^\pm \rangle &: \frac{1}{\sqrt2}(\lvert 01 \rangle \pm e^{i(\varphi_{\textrm{Si}}(t) - \varphi_{\textrm{H}}(t) )} \lvert 10 \rangle ) 
\end{align}

If both nuclear spins are subject to the same noise field (perfectly correlated noise), we can write:

\begin{equation}
    \varphi_{\textrm{Si}}(t) = a_{\textrm{Si}}f(t),~\varphi_{\textrm{H}}(t) = a_{\textrm{H}}f(t)
    \label{eq:same_noise}
\end{equation}

where $f(t)$ describes the noise process. Assuming that the noise is generated by a large number of fluctuators that are weakly coupled to the nuclear spin, the noise has Gaussian statistics. Therefore, we have:

\begin{equation}
    \langle e^{i\delta \varphi (t)} \rangle = e^{-\frac{1}{2}\langle \delta \varphi^2 (t) \rangle}
\end{equation}

Then, the phase accumulation on the entangled state becomes:

\begin{align}
    \langle (\varphi_{\textrm{Si}}(t) \pm \varphi_{\textrm{H}}(t))^2 \rangle & = \langle \varphi_{\textrm{Si}}^2(t)\rangle + \langle \varphi_{\textrm{H}}^2(t)\rangle \pm 2 \langle \varphi_{\textrm{Si}}(t)\varphi_{\textrm{H}}(t)\rangle \notag \\
    & = a_{\textrm{Si}}^2\langle f^2(t) \rangle + a_{\textrm{H}}^2\langle f^2(t) \rangle \pm 2a_{\textrm{Si}}a_{\textrm{H}}\langle f^2(t) \rangle \notag \\
    & \propto \frac{t^2}{(T_{2,~\textrm{Si}}^*)^2} + \frac{t^2}{(T_{2,~\textrm{H}}^*)^2} \pm 2a_{\textrm{Si}}a_{\textrm{H}}\langle f^2(t) \rangle
    \label{eq:noise_expression}
\end{align}

From eq. \ref{eq:same_noise}, we know that
\begin{align}
    \frac{a_{\textrm{Si}}}{a_{\textrm{H}}} &= \frac{T_{2,~\textrm{H}}^*}{T_{2,~\textrm{Si}}^*} \label{eq:math_rep1}\\
    \langle f^2(t) \rangle & = \frac{t^2}{(T_{2,~\textrm{Si}}^*)^2 * a_{\textrm{Si}^2}}
    \label{eq:math_rep2}
\end{align}

Combining eq. \ref{eq:math_rep1}, \ref{eq:math_rep2} into eq. \ref{eq:noise_expression}, we get 
\begin{align}
     \langle (\varphi_{\textrm{Si}}(t) \pm \varphi_{\textrm{H}}(t))^2 \rangle & \propto \frac{t^2}{(T_{2,~\textrm{Si}}^*)^2} + \frac{t^2}{(T_{2,~\textrm{H}}^*)^2} \pm \frac{2t^2}{T_{2,~\textrm{H}}^* \times T_{2,~\textrm{Si}}^*}
     \label{eq:noise_final}
\end{align}
 
For perfectly correlated noise (eq.~\ref{eq:noise_final}), we can use the single qubit coherence times $T_{\textrm{2,~H}}^*$ and $T_{\textrm{2,~Si}}^*$ to estimate Bell state coherence times of $T_2^* = 3.1 (1)$~ms for $\lvert \Phi \rangle$ states, and $T_2^* = 5.7(4)$~ms for $\lvert \Psi \rangle$ states. For uncorrelated noise $\langle \varphi_{\textrm{Si}}(t)\varphi_{\textrm{H}}(t)\rangle=0$, we predict $T_{2}^*=3.9(2)$~ms for either entangled state. In our experiment, we measured different coherence times for $\lvert \Phi^+ \rangle$ ($T_2^* = 2.60(8)$~ms), and $\lvert \Psi^+ \rangle$ ($T_2^* = 3.75(8)$~ms). The different coherence times for the different entangled states suggest that the two nuclear spins are subject to partially correlated noise. The partial noise correlation can be explained by the different quantization axes the two nuclear spins experience since the Zeeman and hyperfine fields are of the same order of magnitude but are not aligned. 

\begin{figure}[htbp]
    \centering
    \includegraphics[]{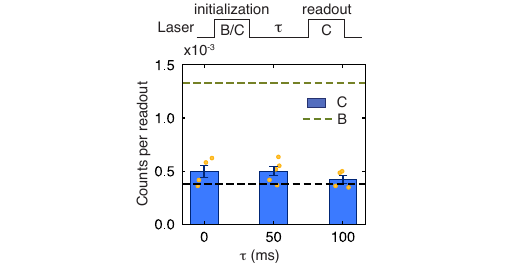}
    \caption{\textbf{Electron spin lifetime measurements} Transition B/C was pumped to initialize the electron spin. By pumping at transition C for initialization, reading out polarization with transition C at different delay times ($\tau$) shows no noticeable polarization decay up to $100$~ms. The black dashed line indicates the dark counts. The green dashed line indicates the expected counts if the spin was initialized by pumping at transition B.}
    \label{fig:SI_electron_T1}
\end{figure}

Finally, we note that electron spin relaxation is unlikely to influence the coherence times of the entangled states. During Bell state measurements, we initialize the electron spin to the $\lvert e_\downarrow \rangle$ state whose lifetime exceeds $100~$ms (Fig.~\ref{fig:SI_electron_T1}), much longer than our sequence time.

\end{document}